\documentclass[iop,apj]{emulateapj}
\usepackage{amsmath,amssymb,color}

\newcommand{\3}{I\hspace{-.1em}I\hspace{-.1em}I}

\newcommand{\Msun}{\ensuremath{\mathrm{M}_\odot}}
\newcommand{\Myr}{\ensuremath{\mathrm{Myr}}}

\begin{document}

\title{Can direct collapse black holes launch gamma-ray bursts and grow to supermassive black holes?}

\author{Tatsuya Matsumoto\altaffilmark{1}, Daisuke Nakauchi\altaffilmark{1,2}, Kunihito Ioka\altaffilmark{3,4}, Alexander Heger\altaffilmark{5,6,7} and Takashi Nakamura\altaffilmark{1}}

\altaffiltext{1}{Department of Physics, Kyoto University, Kyoto 606-8502, Japan}
\altaffiltext{2}{Astronomical Institute, Tohoku University, Aoba, Sendai 980-8578, Japan}
\altaffiltext{3}{Theory Center, Institute of Particle and Nuclear Studies, KEK, Tsukuba 305-0801, Japan}
\altaffiltext{4}{Department of Particle and Nuclear Physics, SOKENDAI (The Graduate University for Advanced Studies), Tsukuba 305-0801, Japan}
\altaffiltext{5}{Monash Centre for Astrophysics, Monash University, Melbourne, Victoria 3800, Australia}
\altaffiltext{6}{University of Minnesota, School of Physics and Astronomy, Minneapolis, MN 55455, USA}
\altaffiltext{7}{Shanghai Jiao-Tong University, Center for Nuclear Astrophysics, Department of Physics and Astronomy, Shanghai 200240, P.~R.~China}

\begin{abstract}
The existence of black holes (BHs) of mass $\sim 10^9\,\Msun$ at
$z\gtrsim6$ is a big puzzle in astrophysics because even optimistic
estimates of the accretion time are insufficient for stellar mass BHs
of $\sim10\,\Msun$ to grow into such supermassive BHs.  A resolution
of this puzzle might be the direct collapse of supermassive stars with
mass $M \sim 10^5\,\Msun$ into massive seed BHs.  We find that if a
jet is launched from the accretion disk around the central BH, the jet
can break out the star because of the structure of the
radiation pressure-dominated envelope.  Such ultra-long
gamma-ray bursts with duration of $\sim 10^4$-$10^6\,\rm{s}$ and
flux of $10^{-11}$-$10^{-8}\,\rm{erg\,s^{-1}\,cm^{-2}}$ could be
detectable by \textit{Swift}.  We estimate an event rate of
$\lesssim1\,\rm{yr}^{-1}$.  The total explosion energy is
$\gtrsim10^{55}$-$10^{56}\,\rm{erg}$.  The resulting negative
feedback delays the growth of the remnant BH by about $70\,\rm{Myr}$
or evacuates the host galaxy completely.

\end{abstract}


\section{Introduction}\label{Introduction}
Gamma-ray bursts~(GRBs) are among the most violent explosions in the
Universe.  They are classified into two populations by the duration of
prompt emission $T_{90}$: short GRBs (SGRBs) with $T_{90}<2\,\rm{s}$
and long GRBs (LGRBs) with $T_{90}>2\ \rm{s}$~\citep{Kouveliotou+93}.
A widely accepted model of LGRBs is the collapsar scenario, in which a
black hole (BH) and accretion disk system is formed after the stellar
collapse, and launches a relativistic jet that breaks out the
progenitor star, producing a GRB \citep{Woosley93,MacFadyen and
  Woosley99}.  Theoretical models identify the typical duration of a
LGRB ($T_{90} \sim 30\,$s) as the free-fall time of the envelope, or
the sound-crossing time of the shocked envelope, of a Wolf-Rayet (WR)
star \citep{Mizuta and Ioka13}.  From observation we know that at
least some LGRBs are accompanied by broad-lined Ic supernovae
\citep[SNe;][]{Woosley and Bloom06,Hjorth and Bloom12}.  This suggests
a tight connection between GRBs and progenitors with stripped envelope
like WR stars.

Recently, some LGRBs have been discovered to show ultra-long duration
of the prompt emission with $\delta t_\gamma \sim 10^4\,$s.  These
have been named ultra-long gamma-ray bursts (ULGRBs)~\citep{Gendre+13,
  Levan+14}.  The ultra-long duration was first predicted in the
context of Population \3 (PopIII) GRBs \citep{Suwa and Ioka11,
  Nagakura+12}, and subsequent studies suggested that a metal-poor
blue supergiant (BSG) collapsar rather than the WR one is more
favorable to explain such bursts \citep{Gendre+13, Kashiyama+13,
  Nakauchi+13}.  Since metal-poor stars may suffer from little mass
loss, metal-poor BSG stars would keep massive hydrogen envelopes until
the pre-collapse phase \citep{Woosley+02}.  Therefore, the accretion
of the massive hydrogen envelope can lead to the long-lasting central
engine activity~\citep{Suwa and Ioka11, Nagakura+12, Quataert and
  Kasen12, Woosley and Heger12, Nakauchi+13}.  Since metal-poor stars
are considered to be the dominant population in the high-$z$ Universe,
ULGRBs might be a dominant population of GRBs in the high-$z$
Universe~\citep[e.g.,][]{deSouza+11}.

On the other hand, the existence of BHs of mass $\sim10^9\,\Msun$ at
$z\gtrsim6$ \citep{Fan06,Mortlock+11,Wu+15} is a great mystery in
astrophysics because the accretion time is not enough to grow the BHs
from the stellar mass BHs of mass $\sim 10\,\Msun$ \citep[e.g.,][]{Haiman13}.
Many attempts are made to solve this problem.  One possible solution
may be the formation from PopIII stars.  Recent numerical simulations
suggest that the mass of PopIII stars reaches up to $10^{2-3}\,\Msun$
\citep{Hirano+14,Susa+14}.  BHs born from these PopIII stars can
barely grow up to supermassive BHs of mass $10^9\,\Msun$ at
$z\gtrsim6$, if successive high mass accretions are maintained.
Feedback effects from accreting BHs, however, decrease the accretion
rate \citep{Alvarez+09}.  Thus it seems difficult for seed BHs from
PopIII stars to form the observed supermassive BHs without resorting
to super-Eddington accretion via such as Bondi and cold accretion
\citep[e.g.,][]{Volonteri:2005fj,Gaspari:2013ah} or efficient mergers
\citep[e.g.,][]{Madau:2001sc}.

An attractive alternative might be the formation of supermassive stars
(SMSs) of mass $\sim 10^5\,\Msun$.  SMSs may be formed in the high
temperature region irradiated by the strong ultra-violet radiation
from nearby galaxies \citep{Omukai01, Bromm and Loeb03, Shang+10,
  Latif+13, Inayoshi+14,Johnson+14}.  When SMSs end their life due to the
exhaustion of their nuclear fuel or the general relativistic (GR)
instability \citep{Chandrasekhar64,Osaki66,Shapiro,Shibata and
  Shapiro02}, they collapse to massive BHs of mass $\sim10^5\,\Msun$,
so-called direct collapse BHs (DCBHs).  If these DCBHs are seeds, the
accretion time might be enough to grow the BHs to $\sim10^9\,\Msun$
at $z\gtrsim6$.

Various violent phenomena from the gravitational collapse of SMSs are
expected such as the energetic neutrino bursts \citep{Fryer and
  Heger11} and SNe with very huge explosion energy of
$\sim10^{55}\ \rm erg$ \citep{Johnson+13,Whalen+13,Chen+14}.  In this
paper, we consider ULGRBs from SMSs as another possibility.  If SMSs
evolve without mass ejection and collapse to form BH--disk systems,
like collapsars, we can expect the launch of the relativistic jet
similar to LGRBs\footnote{When a SMS collapse to a BH, there is some
  possibility that a quasi-star, which is powered by the central BH
  accretion, could form \citep{Begelman+08}.  In this work, however,
  we consider that the central BH launches a jet as ordinary GRBs.
  Previous works \citep{Barkov:2010wm,Czerny:2012gg} did not consider
  the jet propagation in the SMS, so that the observed quantities such
  as the luminosity and duration are unable to obtain.  }.  Since SMSs
are larger in radius than PopIII first stars of mass
$10$-$1000\,\Msun$ or BSGs, we expect that the duration of ULGRBs
is even longer than that of the observed ULGRBs.  The
detection of such ULGRBs would enable us to observe the very moment of
the birth of first quasars to probe the high-$z$ Universe.  Such
energetic explosions might prevent the subsequent gas accretion and
the growth of the DCBHs.  Thus, it is worth evaluating the details of
relativistic jet explosions from supermassive collapsars and their
observational signatures.

This paper is organized as follows: In \S \ref{Progenitor Model}, we
show the pre-collapse stellar models of SMSs, which evolve from zero
age main sequence (ZAMS) stars.  In \S \ref{GRB associated with the
  SMSs}, we describe the method to calculate the jet dynamics in the
SMS envelope.  In \S \ref{Results}, we show that the jet can break out
the SMS.  We also discuss the observational signatures and the
detectability of GRBs from supermassive collapsars.  In \S
\ref{Discussion}, we discuss the other progenitor model in which the
SMS is accreting mass and collapses through the GR instability. Then
we estimate the event rate of GRBs from SMSs, and discuss the effects
of GRBs on their environment.  A summary and our conclusions are given
in \S \ref{Summary}.  Throughout this paper, we consider the
$\Lambda${CDM} cosmology and adopt the cosmological parameters as :
$H_0=67.8\,\rm{km\ s^{-1}\,Mpc^{-1}}$, $\Omega_{\rm{m}}=0.308$ and
$\Omega_{\Lambda}=0.692$~\citep{Planck15}.

\section{Progenitor Model}\label{Progenitor Model}

Theoretical studies of the primordial star formation have suggested
that SMSs with $\gtrsim 10^5\,\Msun$ can be formed under the hot
environments of first galaxies~\citep{Omukai01, Bromm and Loeb03,
  Shang+10, Latif+13, Inayoshi+14,Johnson+14}.  Whereas molecular
hydrogen is the primary coolant in primordial star-forming clouds, its
formation is prevented under strong ultra-violet radiation from nearby
galaxies, so that the temperature is kept at $\sim 10^4\,\rm K$ in
such a cloud due to the hydrogen atomic cooling.  The accretion rate
onto a protostar $\dot{M}$ can be evaluated by dividing the Jeans mass
$M_{\rm J}$ of the star-forming cloud by the free-fall time $t_{\rm
  ff}$:
\begin{equation}
\dot{M} \sim M_{\rm J} / t_{\rm ff} \sim c_{\rm s}^3 / G \sim 0.51
\left(\frac{T}{10^4\ {\rm K}}\right)^{3/2}\,\Msun \ {\rm yr}^{-1},
\label{eq:m_acc}
\end{equation}
where $c_{\rm{s}}$ and $T$ are the sound velocity and the temperature
of the cloud, and $G$ is the gravitational constant, respectively.  We
also assume that the cloud is composed of only hydrogen, for
simplicity.  Thus, the mass accretion rate onto the protostar formed
in the central region of the star-forming cloud can be as high as
$\sim0.1$-$1\,\Msun\,\rm{yr}^{-1}$ under such hot environments.
The protostar can grow up to a SMS with $\gtrsim 10^5\,\Msun$ within
its lifetime of $\sim1\,\Myr$ through such a high mass accretion rate.

When the mass accretion onto the protostar stops because of radiation
feedback, or for other reasons, the protostar contracts onto the ZAMS
in a Kelvin-Helmholtz timescale.  Since the SMSs are almost fully
convective and the Kelvin-Helmholtz timescale 
 is short compared to the lifetime of the star, the resulting final stellar structure should be the
same as long as accretion stops not too close to hydrogen depletion.
After the SMS runs through its nuclear burning phases\footnote{The
  advanced phases may be accelerated as part of the collapse.}, it
collapses to a BH similar to a massive star\footnote{Even a phase with
  a hot proto-BH due to neutrino trapping as observed by
  \citet{Fryer+01} may nor occur.}  When the SMS collapses to a BH, it
may produce a GRB by launching a relativistic jet provided there is
enough angular momentum.  Thus, we adopt this pre-collapse SMS for our
progenitor model.

\begin{figure}[!t]
\centering
\includegraphics[scale=0.3,angle=270]{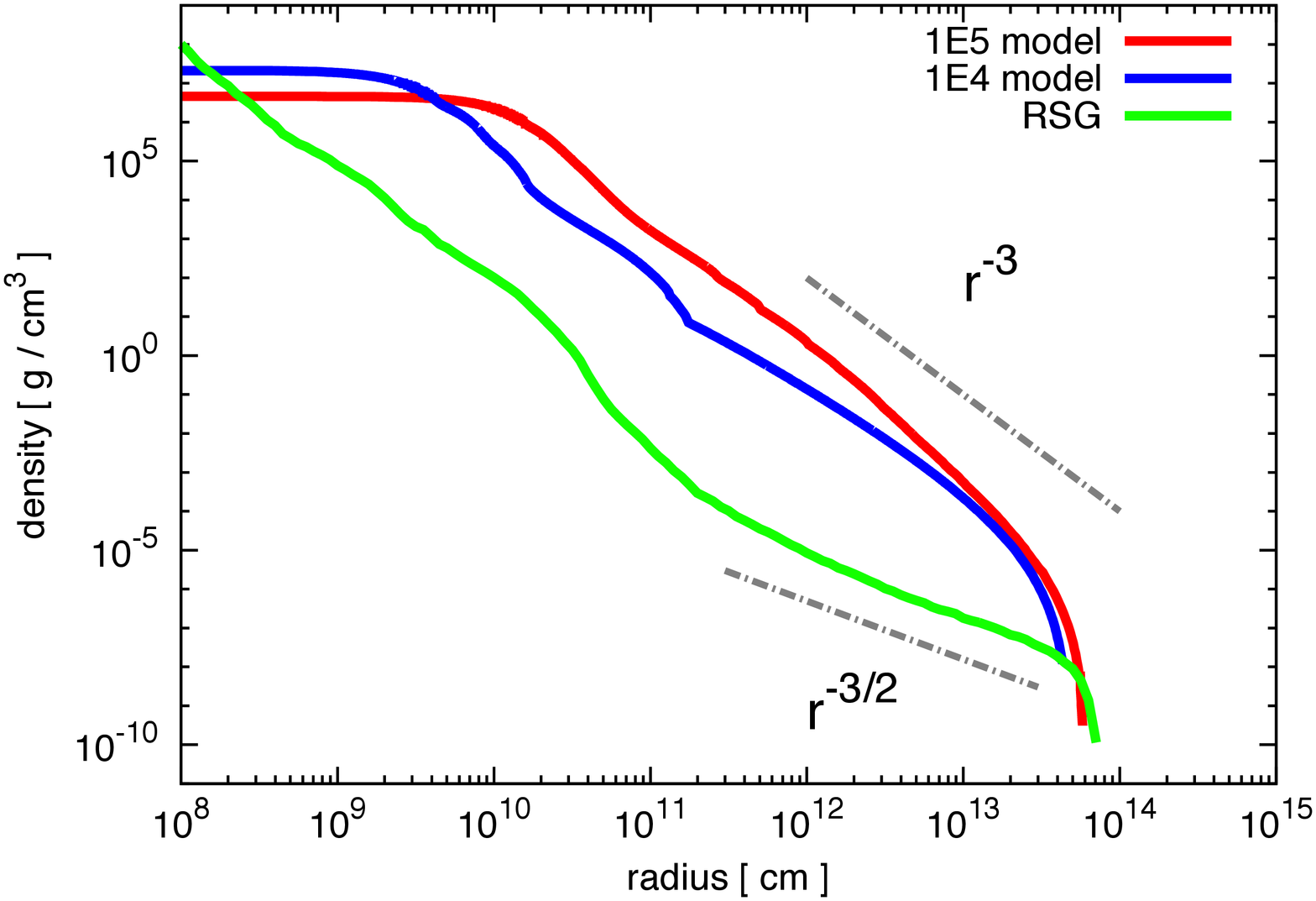}
\caption{Density profiles of SMSs in their pre-collapse phase.  The
  \emph{red curve} and \emph{blue curve} correspond to the 1E5 model
  and the 1E4 model, respectively.  These profiles are obtained by
  calculating the metal-free $10^5\,\Msun$ and $10^4\,\Msun$ ZAMS stars
  until the nuclear fuels are exhausted in the center.  The
  \emph{green curve} shows the density profile of a RSG (pre-supernova
  structure of Model \texttt{s15} from \citealt{Woosley+02}).  These
  profiles show that SMSs have similar radii with a RSG but steeper
  density profile than that of a RSG.  The \emph{gray dash-dotted lines}
  show the slope of the power-law density profiles of
  $\rho\propto{r^{-3}}$ and $\rho\propto{r^{-3/2}}$ as references.}
\label{density profile image}
\end{figure}

In this paper, we focus on a supermassive progenitor with a ZAMS mass
of $10^5\,\Msun$, but for comparison also consider a progenitor which
has the ZAMS mass of $10^4\,\Msun$ \citep{Fryer and Heger11}.
  We call the former model as ``1E5
model'' and the latter model as ``1E4 model'', respectively.
  The density profiles of these pre-collapse SMSs are shown
in Figure~\ref{density profile image}.  We can see that they have very
large radii of $R_\ast \sim 10^{14}\,\rm cm$, which are as large as
those of present-day RSGs (green curve in Fig.~\ref{density profile
  image}).  Note that the envelopes of the SMS models have steeper
density profiles ($\rho \propto r^{-3}$) than that of the RSG model
($\rho \propto r^{-3/2}$).  This is because radiation pressure
dominates in the SMS envelopes while gas pressure dominates in the RSG
envelope (see Section \ref{breakout}).

It is not trivial whether the relativistic jet can break out the
envelope of the supermassive progenitors successfully since it can not
break out the large envelope of RSGs~\citep{Matzner03, Suwa and
  Ioka11, Nakauchi+12}.  \cite{Suwa and Ioka11, Nagakura+12} studied
relativistic jet explosions from massive PopIII stars
($M_*\simeq10^3\,\Msun$), and found that the jet breakout is possible
despite the large radius of $R_\ast \sim 10^{13}\,\rm cm$. They
attributed this to the long duration of the mass accretion of the
massive envelope.  In this paper, we consider supermassive PopIII
stars which have $M_*\simeq10^4$-$10^5\,\Msun$ and
$R_\ast\sim10^{14}\,\rm cm$ to determine whether the jet can break out
of the envelope.  These stars have $10$ times larger radii compared to
the massive PopIII stars considered by \cite{Suwa and Ioka11,
  Nagakura+12} so that the jet might not break out the envelope.

\section{Numerical Method}\label{GRB associated with the SMSs}

\subsection{Jet Model}\label{Jet model}
We adopt a simple jet model in which the jet luminosity varies with
time and depends on the mass accretion rate onto the BH~\citep{Suwa
  and Ioka11}.  The SMS begins to collapse, when its nuclear fuel is
exhausted in the center, and forms a massive BH.  We evaluate the mass
accretion rate onto the central BH using the free-fall timescale as
typical timescale.  For each mass shell at radius $r$ and mass
coordinate $M_r$, the free-fall timescale can be calculated from
\begin{eqnarray}
t_{\rm{ff}}(r)=\sqrt{\frac{3\pi}{32G\bar{\rho}}}=\frac{\pi}{2}\sqrt{\frac{r^3}{2GM_r}},
\label{free fall time}
\end{eqnarray}
where $\bar{\rho}=M_{r}/\frac{4\pi}{3}r^3$ is the mean density within
$r$.  The mass accretion rate is calculated by~\citep{Woosley and
  Heger12,Woosley and Heger15}
\begin{eqnarray}
  \dot{M}=\frac{dM_r}{dt_{\rm{ff}}}=\frac{{dM_r}/{dr}}{{dt_{\rm{ff}}}/{dr}}=
  \frac{2M_r}{t_{\rm{ff}}(r)}\left(\frac{\rho}{\bar\rho-\rho}\right).
\label{mass accretion rate}
\end{eqnarray}

A relativistic jet will be launched from the central engine, which is
composed of a BH and an accretion disk.  The mechanisms of launching
relativistic outflows are still uncertain, though several processes
have been proposed such as MHD mechanism and neutrino-antineutrino
annihilation mechanism~\cite[e.g.,][]{Blandford and
  Znajek77,Popham+99}.  \cite{Suwa and Ioka11} studied the jet-driven
explosions from massive PopIII stars and found that the MHD mechanism
works long enough for the jet to break out the envelope, whereas the
neutrino mechanism does not.  Therefore, we assume the MHD mechanism
hereafter.  In the MHD process, the jet luminosity can be modeled
according to \citep{Komissarov and Barkov10}
\begin{eqnarray}
L_{\rm{j}}(t)=\eta_{\rm{j}}\dot{M}(t)c^2,
\label{jet luminosity}
\end{eqnarray}
where $\eta_{\rm{j}}$ is the efficiency parameter.  For simplicity we
assume that $\eta_{\rm{j}}$ is constant. We can take into account the
effect of the progenitor rotation by adjusting the parameter value
$\rm{\eta_{\rm{j}}}$.  When a sufficiently fast-rotating progenitor
collapses, it can form a BH and accretion disk system.  The remainder
of the star then falls onto the central BH in the accretion time
$t_{\rm{acc}}$, which can be related to the free-fall time as
$t_{\rm{acc}}\propto \frac{1}{\alpha}t_{\rm{ff}}$, where $\alpha$ is
the viscosity parameter \citep{Shakura and Sunyaev73, Kumar+08}.
Thus, the factor $\alpha$ can be absorbed into the effective value of
$\eta_{\rm{j}}$.  We use $\eta_{\rm{j}}=6.2\times10^{-4}$, which
is calibrated to reproduce the total jet energy of typical LGRBs of
$E_{\rm{tot}}=\int{\eta_{\rm{j}}\dot{M}c^2}dt=10^{52}\,\rm{erg}$ for a
WR progenitor model~\citep{Suwa and Ioka11}.

Throughout this section, the time $t$ is measured in the central
engine frame. We set $t=0$ as the time when the black hole is
formed. Since the formation mechanism of a relativistic jet is still
under debates, the time of jet formation, $t_{\rm{in}}$, is also
uncertain and should be a parameter of this study.  Following the
previous studies, we assume that the jet is formed when the mass of BH
reaches $3\,\Msun$: $t_{\rm{in}} = t_{\rm{ff}}(r_{\rm{in}})$, where
the enclosed mass within the radius $r_{\rm{in}}$ is $3\,\Msun$, i.e.,
$M_{r_{\rm{in}}}=3\,\Msun$~\citep{Suwa and Ioka11}.  The choice of
$t_{\rm{in}}$~(or $M_{r_{\rm{in}}}$), however, has little influence on
the jet dynamics in the envelope as long as $t_{\rm{in}}$ is much
smaller than the jet break out time.  In fact, we find that the jet
breakout time is within a factor of two, when the jet formation time
is changed as $M_{r_{\rm{in}}}=30\,\Msun$, $300\,\Msun$, and
$3,\!000\,\Msun$ in the 1E5 model.

\subsection{Jet Propagation in the SMS Envelope}\label{Jet Propagation in the SMS}
Here, we describe the jet dynamics in the progenitor envelope
following the prescription in~\cite{Matzner03, Suwa and Ioka11,
  Bromberg+11,Nakauchi+12}.  A relativistic jet launched from the
central engine collides with the stellar matter and forms the shocked
region at the jet head.  The jet head is composed of the forward shock
which sweeps the stellar matter and the reverse shock which
decelerates the jet matter.  Both shocked matter are divided by the
contact discontinuity.  Then the velocity of the jet head can be
calculated from the pressure balance at the contact discontinuity as
\begin{eqnarray}
\rho_{\rm{j}}c^2h_{\rm{j}}\Gamma_{\rm{j}}^2\Gamma_{\rm{h}}^2(\beta_{\rm{j}}-\beta_{\rm{h}})^2+P_{\rm{j}}=\rho_{\rm{a}}c^2h_{\rm{a}}\Gamma_{\rm{h}}^2\beta_{\rm{h}}^2+P_{\rm{a}},
\label{jet head}
\end{eqnarray}
where, $\rho$, $h$, and $P$ represent the mass density, specific
enthalpy, and pressure measured in the fluid rest frame, respectively,
and $\beta$ and $\Gamma=(1-\beta^2)^{-1/2}$ are the velocity
normalized by the speed of light $c$ and the Lorentz factor,
respectively.  The subscripts ``h'', ``j'', and ``a'' indicate that
the quantity is measured in the rest frame of the jet head, the jet,
and the ambient stellar medium, respectively.  Since the stellar
medium is composed of the non-relativistic matter, we can neglect its
pressure against the rest mass energy density, i.e., $P_{\rm{a}} \ll
\rho_{\rm{a}}c^2$, and this also leads to $h_{\rm{a}}\simeq1$.  In the
left hand side of Equation (\ref{jet head}), we can also neglect the
jet pressure, since the jet is ultra-relativistic
($\beta_{\rm{j}}\simeq1$ and $\Gamma_{\rm{j}}\gg1$).

Using the above approximations, the velocity of the jet head is given
by
\begin{eqnarray}
\beta_{\rm{h}}&\simeq&\frac{1}{1+\tilde{L}^{-\frac{1}{2}}},
\label{jet head velocity}
\end{eqnarray}
where $\tilde{L}=\rho_{\rm{j}}h_{\rm{j}}\Gamma_{\rm{j}}^2/\rho_{\rm{a}}$ is a parameter which determines the jet dynamics.
Using the collimation-corrected jet luminosity given by
\begin{eqnarray}
L_{\rm{j}} = \rho_{\rm{j}}c^3h_{\rm{j}}\Gamma_{\rm{j}}^2\beta_{\rm{j}}\Sigma_{\rm{h}},
\end{eqnarray}
$\tilde{L}$ is given by
\begin{eqnarray}
\tilde{L}\simeq\frac{{L}_{\rm{j}}}{\rho_{\rm{a}}c^3\Sigma_{\rm{h}}},
\label{jet parameter}
\end{eqnarray}
where $\Sigma_{\rm{h}}$ is the cross section of the jet head.  Thus,
$\tilde{L}$ is the ratio of the luminosity of the jet to that of the
ambient stellar medium rest energy flux.  Here, we assume that the
opening angle of the jet is constant with $\theta=5^\circ$.  This is
the typical value obtained from the afterglow observations of
LGRBs~\citep{Frail+01}.  Then the cross section can be given by
$\Sigma_{\rm{h}}=\pi(r_{\rm{h}}\theta)^2$, where
$r_{\rm{h}}(t)=\int{c\beta_{\rm{h}}}dt$ is the radius of the jet head.

As long as the velocity of the jet head is
non-relativistic~($\beta_{\rm h} < 1$), the shocked jet head can
expand sideways to form a cocoon structure surrounding the jet.  Since
the temperature of the cocoon is high, it is
radiation-pressure-dominated.  As long as the sound crossing time in
the cocoon is shorter than the dynamical time of the jet head, we can
neglect the inner structure of the cocoon and assume that the cocoon
is uniform.  Hereafter, we consider the one-zone model for the lateral
expansion of the cocoon.  The cocoon is overpressured with respect to
the ambient stellar medium so that it expands laterally.  By
considering pressure balance at the surface of the cocoon, the lateral
expansion velocity of the cocoon is calculated by \citep{Begelman and
  Cioffi89}
\begin{eqnarray}
\beta_{\rm{c}}&=&\sqrt{\frac{P_{\rm{c}}}{\overline{\rho}_{\rm{a}}c^2}},
\label{cocoon velocity}
\end{eqnarray}
where $P_{\rm{c}}$ is the pressure in the cocoon and
$\bar{\rho}_{\rm{a}}(r_{\rm{h}}) = M_{r_{\rm{h}}}/(4 \pi r_{\rm{h}}^3
/3)$ is the mean density of the progenitor star within the radius
$r_{\rm{h}}$.

Since the cocoon matter is radiation-pressure-dominated, the pressure
is given by $P_{\rm{c}}=E_{\rm{c}}/3V_{\rm{c}}$, where $E_{\rm{c}}$
and $V_{\rm{c}}$ represent the total energy and volume in the cocoon,
respectively.  Because the cocoon energy is supplied from the jet
head, it is given by
\begin{eqnarray}
E_{\rm{c}}(t)=\eta_{\rm{c}}\int_{t_{\rm{in}}}^{t-\frac{r_{\rm{h}}(t)}{c}}{L}_{\rm{j}}(t^\prime)dt^\prime,
\label{cocoon energy}
\end{eqnarray}
where $\eta_{\rm{c}}$ indicates the fraction of the jet luminosity
streaming into the cocoon.  Throughout the paper, we set
$\eta_{\rm{c}}=1$, since the velocity of the jet head is
non-relativistic for most of the time within the progenitor envelope.
In Equation (\ref{cocoon energy}), the upper limit of the integral
indicates that at $t$, the jet head receives the luminosity which is
produced at $t - r_{\rm{h}}(t)/c$ at the central engine.  For the
cocoon volume, we assume that the cocoon has a conical shape with the
height of $r_{\rm{h}}(t)$ and the base radius of ${r}_{\rm{c}}(t)$, so
that it is given by
\begin{eqnarray}
V_{\rm{c}}(t)=\frac{1}{3}\pi{r}_{\rm{c}}^2(t)r_{\rm{h}}(t),
\label{cocoon volume}
\end{eqnarray}
where $r_{\rm{c}}(t)$ is the lateral distance of the cocoon surface
from the jet axis given by $r_{\rm{c}}(t)=\int{c\beta_{\rm{c}}}dt$.

Substituting Equations (\ref{jet luminosity}) and (\ref{jet parameter})
and the definition of $\Sigma_{\rm{h}}$ into Equation (\ref{jet head
  velocity}), the jet head velocity is given by
\begin{eqnarray}
\beta_{\rm{h}}&\simeq&\left[1+\left(\frac{\pi c
    \theta^2\rho_{\rm{a}}(r_{\rm{h}})r_{\rm{h}}^2}{\eta_{\rm{j}}\dot{M}(t)}\right)^{1/2}\right]^{-1}.
\label{jet head velocity2}
\end{eqnarray}
Substituting Equations (\ref{cocoon energy}) and (\ref{cocoon volume})
into (\ref{cocoon velocity}), the lateral expansion velocity of the
cocoon is given by
\begin{eqnarray}
\beta_{\rm{c}}&\simeq&\frac{2r_{\rm{h}}}{cr_{\rm{c}}}\left[\frac{\eta_{\rm{c}}\int_{t_{\rm{in}}}^{t-\frac{r_{\rm{h}}}{c}}L_{\rm{j}}(t^\prime)dt^\prime}{3(M_{r_{\rm{h}}}-M_{r_{\rm{in}}})}\right]^{1/2}.
\label{cocoon velocity2}
\end{eqnarray}

\section{ULGRBs from Supermassive Collapsars}\label{Results}

\subsection{Successful Breakout of Supermassive Collapsar Jets}\label{breakout}

\begin{figure}[!t]
\centering
\includegraphics[scale=0.3,angle=270]{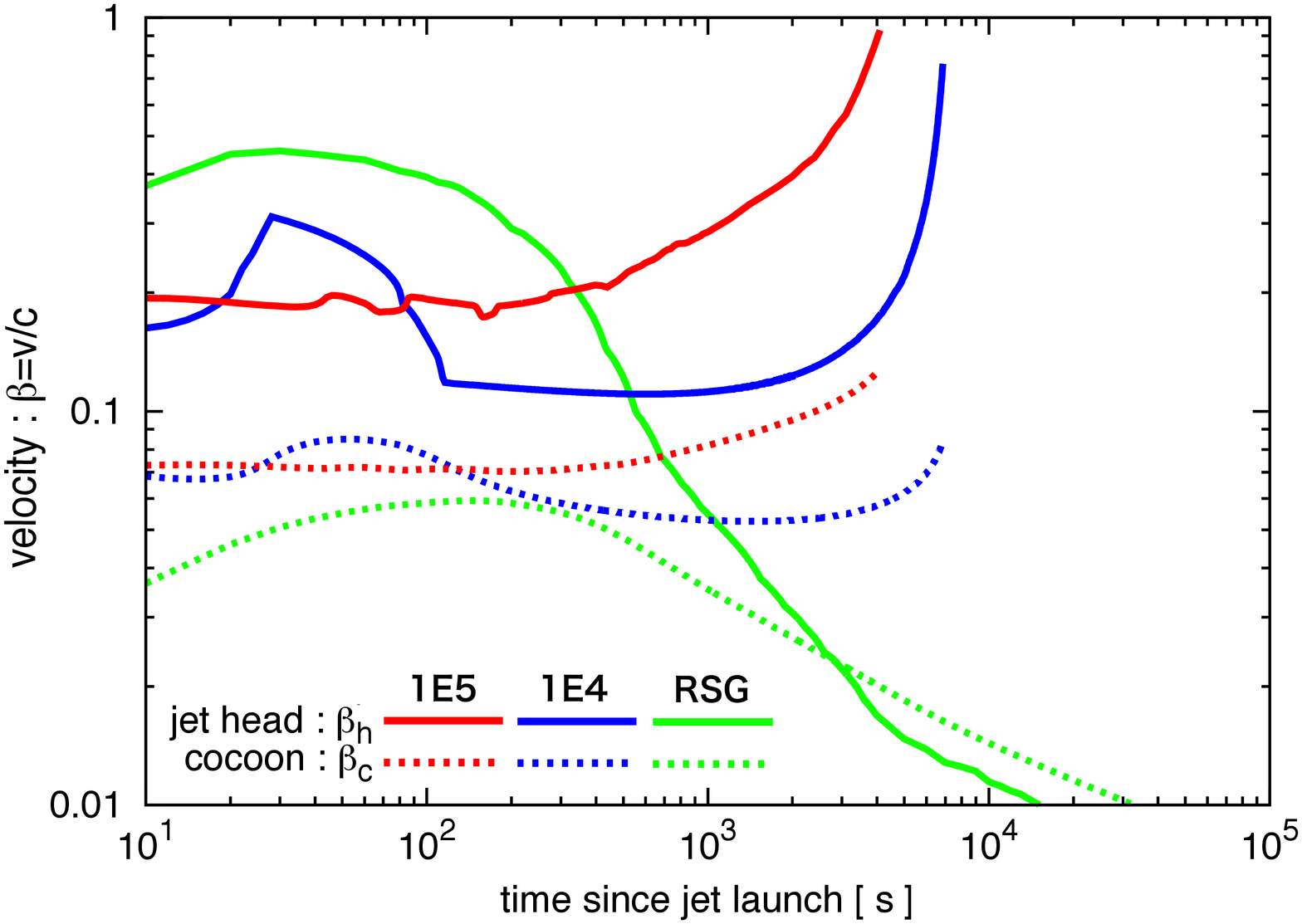}
\caption{Time evolution of the velocities of the jet head and the
  lateral expansion of the cocoon surface.  \emph{Red}, \emph{blue} and \emph{green
    lines} correspond to the 1E5 model, the 1E4 model and the RSG,
  respectively.  The horizontal axis shows the time from jet
  formation.  The vertical axis shows the velocity divided by
  the speed of light.  The \emph{solid} and \emph{dashed curves}
  correspond to the velocities of the jet head $\beta_{\rm{h}}$ and
  the cocoon surface $\beta_{\rm{c}}$, respectively.  For the 1E5
  model and the 1E4 model, the jet head velocity is always larger than
  that of the cocoon edge so that the supermassive collapsar jet can
  break out the envelope.  For $t\gtrsim 500\,\rm{s}$ and
  $t\gtrsim2,\!000\,\rm{s}$, the jet head is accelerated drastically
  in the 1E5 and 1E4 models, respectively, because the envelope
  density decreases more steeply than $\propto r^{-3}$.  On the other hand,
  in the RSG, the jet head velocity is overtaken by that of the cocoon
  at $\sim 3,\!000\,$s.  }
\label{velocity evolution image}
\end{figure}

In Figure \ref{velocity evolution image}, we show the time evolution
of the velocities of the jet head and the lateral expansion of the
cocoon surface with solid and dashed curves, respectively.  Each color
corresponds to the 1E5 model (red), the 1E4 model
(blue), and the RSG model (green), respectively.  They
are calculated from Equations (\ref{jet head velocity2}) and
(\ref{cocoon velocity2}).  The horizontal axis shows the time
from jet formation: $t-t_{\rm in}$.  The vertical axis shows
the velocity divided by the speed of light.

We can see that for the models 1E4 and 1E5, the velocity of the jet
head is always larger than that of the cocoon surface.  On the other
hand, for the RSG model, the velocity of the cocoon exceeds that of
the jet head at $\gtrsim 3,\!000$ s.  In the latter case, the radius
of the jet head is comparable to the lateral size of the cocoon
surface so that they can reach the stellar surface almost at the same
time. This looks like a spherical explosion rather than a collimated
explosion. On the other
hand, in the former case, the collimated jet breaks out the progenitor
surface since the jet head reaches the surface much earlier than the
cocoon.

\begin{figure}[!t]
\centering
\includegraphics[scale=0.3,angle=270]{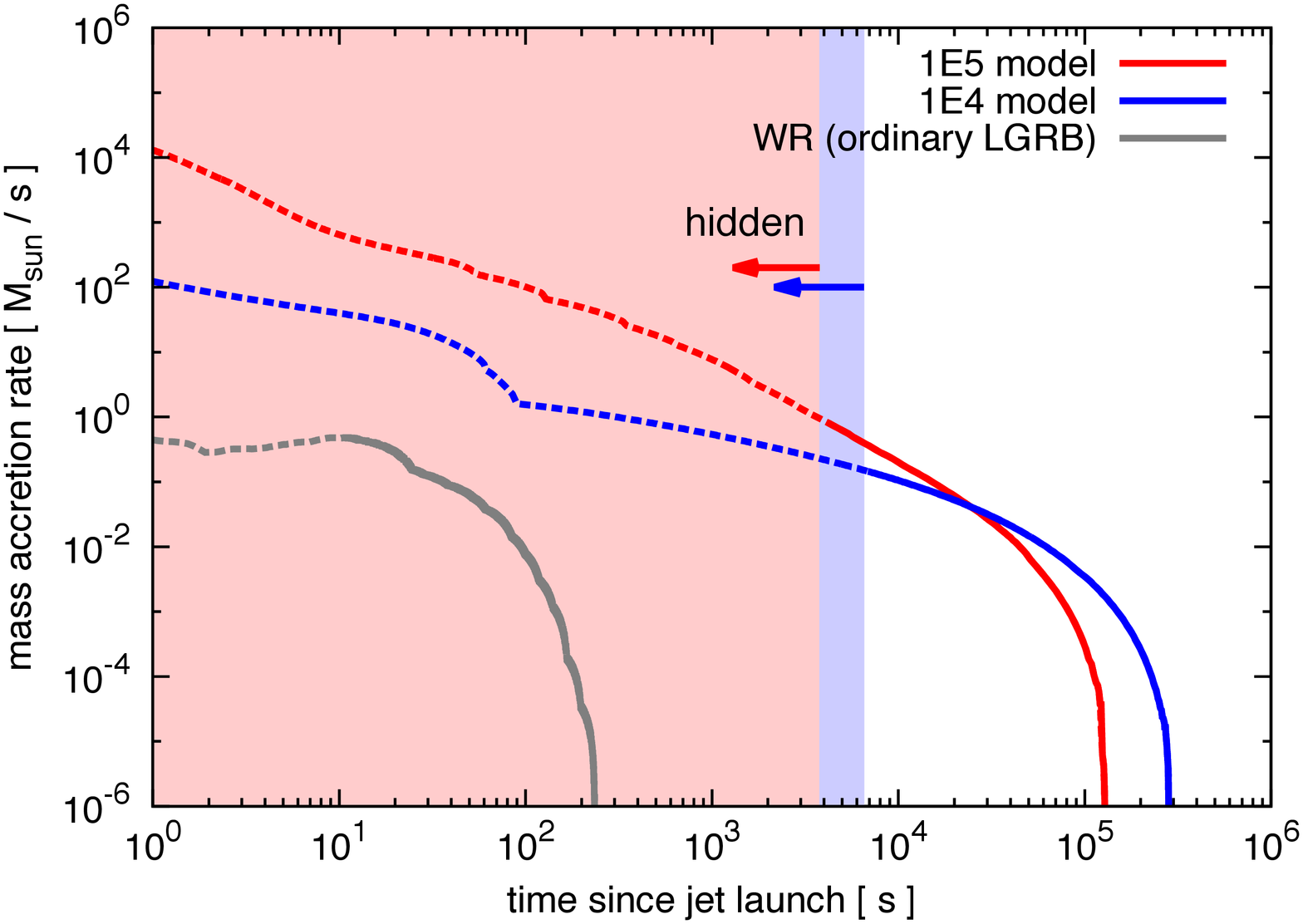}
\caption{Time evolution of the mass accretion rate onto the BH.  The
  horizontal axis is the time from jet formation.  The \emph{red} and
  the \emph{blue lines} corresponds to the accretion rate obtained
  from the 1E5 model and the 1E4 model, respectively. Whole in each
  color, \emph{dashed} and \emph{solid} lines correspond to the time
  before and after the jet breakout, respectively. The \emph{shaded
    region} in each color means that the jet head propagates in the
  progenitor's interior and the jet is ``hidden'' from the
  observers. The accretion rate obtained from the WR model
  corresponding to LGRBs is shown by the \emph{gray curve} as a
  reference. We can see that supermassive collapsar jets can lead to
  ULGRBs with the duration of $T_{90} \sim 10^4\,\rm{s}$ up to $\gtrsim 10^5\,\rm s$. 
  Note that the light curves decline steeply after the breakout, so that the observed duration $T_{90}$ is approximately equal to the breakout time. }
\label{mass accretion rate image}
\end{figure}

In Figure \ref{mass accretion rate image}, we show the time evolution of the mass accretion rate onto the BH.
The horizontal axis represents the time from jet formation.
The red   and  blue curves correspond to the accretion rate obtained from the 1E5 model and the 1E4 model, respectively.
While on each curve, dashed and solid regions correspond to the time before and after the jet breakout, respectively.
We also show the accretion rate obtained from the WR model (i.e., LGRB) with the gray curve as a reference.
We can see that supermassive collapsar jets can lead to ULGRBs with the duration of $\gtrsim 10^5\ \rm s$, owing to the accretion of the massive envelope.

While the SMS models have as large radii as present-day RSGs, we find that the relativistic jet can break out the progenitor envelope successfully.
This can be attributed to the difference in the slope of the density profile.
At the outer envelope of a polytropic star with the polytropic index $n$, the density profile can be approximated as $\rho_{\rm{a}}(r) \propto (R_\ast /r-1)^n \sim r^{-n}$~\citep{Matzner and McKee99}.
Since SMSs have radiation-pressure-dominated convective envelopes, the index can be approximated as $n=3$  \citep{Shapiro,Kippenhahn}.
On the other hand, RSGs have gas-pressure-dominated convective envelopes  so that the index is  $n=1.5$.
Thus, the SMS models have steeper density profiles than the RSG models.

When the velocity of the jet head is non-relativistic, $\tilde{L}$ is less than unity, so that the velocity can be approximately evaluated as
\begin{eqnarray}
\beta_{\rm{h}}(t) \sim \tilde{L}(t)^{1/2} \propto {\dot{M}}(t)^{\frac{1}{2}} r_{\rm{h}}(t)^{\frac{n-2}{2}}.
\label{jet head velocity3}
\end{eqnarray}
Furthermore, by using that $\dot{M}$ is approximated as $\dot{M}(t) \propto t^{\frac{3-2n}{3}}$ in the density profile of $\rho_{\rm{a}}(r) \propto r^{-n}$ \citep{Suwa and Ioka11}, and that $r_{\rm{h}}$ is evaluated roughly as $r_{\rm{h}} \propto \beta_{\rm{h}}t$, we obtain
\begin{eqnarray}
\beta_{\rm{h}}(t) \propto r_{\rm{h}}(t)^{\frac{n-3}{9-2n}},
\label{jet head velocity4}
\end{eqnarray}
from Equation (\ref{jet head velocity3}). 
Therefore the jet head is accelerated in the outer region of the SMS envelopes where the density profile decreases more steeply than $\propto r^{-3}$. On the other hand, the jet head is decelerated in the RSG envelopes.
From Figure \ref{velocity evolution image}, we can see that the jet heads are accelerated drastically at $t \gtrsim 500\ \rm{s}$ and $t \gtrsim 2,\!000\ \rm{s}$ for the 1E5 and 1E4 models, respectively, when the jet heads enter the regions where the envelope density decreases more steeply than $\propto r^{-3}$.
On the other hand, for RSGs, the velocity of the jet head is decelerated in the envelope  so that it takes much longer time to break out the envelope.
In this case, the lateral size of the cocoon becomes comparable to the radius of the jet head.
This is a spherical explosion but not a collimated GRB.
Thus, RSGs cannot be the progenitor of LGRBs as shown by ~\cite{Matzner03,Suwa and Ioka11, Nakauchi+12}.

It is possible that a disk wind flows out isotropically from the accretion disk and changes the envelope structure.
In this case, the disk wind and the deformed envelope may affect the jet propagation.
Unless the wind velocity $v_{\rm{w}}$ is larger than the jet head and cocoon velocity, however, this effect can be ignored.
The wind velocity is evaluated from the wind energy as $E_{\rm{w}}\sim{M}_{\rm{w}}v_{\rm{w}}^2/2$, where $M_{\rm{w}}$ is the mass of the wind component.
The wind energy is also given by $E_{\rm{w}}\sim\int\eta_{\rm{w}}\dot{M}c^2dt$, where $\eta_{\rm{w}}$ is the efficiency parameter, as defined for the jet luminosity in Equation (\ref{jet luminosity}).
On the other hand, the cocoon velocity is evaluated by $E_{\rm{c}}\sim{M_{\rm{c}}}v_{\rm{c}}^2/2\sim\int\eta_{\rm{j}}\dot{M}c^2dt$.
From these equations, we obtain the ratio of the cocoon velocity to the wind velocity as $v_{\rm{c}}/v_{\rm{w}}\sim(M_{\rm{w}}\eta_{\rm{j}}/M_{\rm{c}}\eta_{\rm{w}})^{1/2}$.
In our work, the jet efficiency parameter is $\eta_{\rm{j}}\sim6\times10^{-4}$.
As the wind flows out isotropically while the cocoon expands around the jet, their masses are related as $M_{\rm{c}}\sim{M_{\rm{w}}}\theta^2/2$, where $\theta\sim0.1$ is the jet opening angle.
Then, we have $v_{\rm{c}}/v_{\rm{w}}\sim0.3\eta_{\rm{w}}^{-1/2}$.
Thus, when the disk wind flows out very efficiently ($\eta_{\rm{w}}\gtrsim0.1$), we should consider the effect of the isotropic wind outflow on the envelope structure and the jet propagation.
The recent numerical simulations of super-Eddington accretion disks suggest the value of $\eta_{\rm{w}}\sim0.01$ \citep{Jiang+14}\footnote{\cite{Sadowski+14} also gives the efficiency  $\sim0.3$, although this value includes the efficiency from the accretion energy to the radiation and magnetic energy. }, which yields $v_{\rm{c}}\sim3v_{\rm{w}}$.
It should be noted that the accretion rates in their simulations are $\dot{M}\sim100\,L_{\rm{Edd}}$, while in our situation, the mass accretion rates amount to  $\dot{M}\gtrsim10^{10}L_{\rm{Edd}}$.
We need numerical calculations in order to study the disk wind in our case.

\subsection{Prompt emission}\label{Prompt emission}
Once the relativistic jet breaks out the progenitor's envelope, it can
contribute to the prompt high-energy emission, like LGRBs.  Here, we
evaluate the observational signatures and the detectability of the
prompt emission from the supermassive collapsar jets.  Since the
emission mechanisms of the prompt emission of GRBs are still under
debates, however, following \cite{Nakauchi+12}, we evaluate them by
applying simple empirical relations to supermassive collapsar jets.

First of all, we describe our model for the prompt emission.  We
assume that once the jet breaks out the surface at $t_{\rm{b}}$, the
relativistic jet can contribute to the gamma-ray emission by using a
fraction $\epsilon_{\gamma}$ of its energy.  Thus, the
collimation-corrected gamma-ray luminosity is given by
$L_{\gamma}(t)=\epsilon_{\gamma}L_{\rm{j}}(t)$.  We also assume that
the high energy emission lasts until all the matter in the envelope
has accreted onto the central BH at $t_{\rm{ff},*}=t_{\rm{ff}}(R_*)$.
Hence, the duration of the prompt emission can be evaluated by
$t_{\rm{ff},*} - t_{\rm{b}}$.  Hereafter, we adopt $\theta=5^\circ$,
$\epsilon_{\gamma}=0.1$ and $\eta_{\rm{j}}=6.2\times10^{-4}$ as our
fiducial values.  Then the isotropic luminosity of the prompt emission
is given by
$L_{\gamma,\rm{iso}}(t)=\frac{2}{\theta^2}\epsilon_{\gamma}\eta_{\rm{j}}\dot{M}(t)c^2$.
In this model, the isotropic radiated energy of the prompt emission
$E_{\gamma,\rm{iso}}$ and the peak luminosity $L_{\rm{p}}$ are
estimated by
\begin{eqnarray}
E_{\gamma,\rm{iso}}&=&\frac{2}{\theta^2}\int_{t_{\rm{b}}}^{t_{\rm{ff},*}}L_{\gamma}(t)dt=\frac{2}{\theta^2}{\epsilon_{\gamma}\eta_{\rm{j}}c^2}\int_{t_{\rm{b}}}^{t_{\rm{ff},*}}\dot{M}(t)dt\nonumber\\
&=&2.9\times10^{52}\biggl(\frac{\int_{t_{\rm{b}}}^{t_{\rm{ff},*}}\dot{M}dt}{1\,\Msun}\biggr)\,\rm{ergs},
\label{isotropic gamma ray energy}
\end{eqnarray}
and
\begin{eqnarray}
L_{\rm{p}}&=&L_{\gamma,\rm{iso}}(t=t_{\rm{b}})=\frac{2}{\theta^2}\epsilon_{\gamma}\eta_{\rm{j}}c^2\dot{M}(t=t_{\rm{b}})\nonumber\\
&=&2.9\times10^{52}\biggl(\frac{\dot{M}(t=t_{\rm{b}})}{1\,\Msun\,\,\rm{s^{-1}}}\biggr)\,\rm{ergs\,\,s^{-1}}.
\label{luminosity peak}
\end{eqnarray}
As long as the luminosity is proportional to the mass accretion rate, the luminosity decreases monotonically after the breakout.
Therefore, the luminosity peaks at the breakout.

Next, we evaluate the spectral peak energy of the prompt emission in the central-engine frame $E_{\rm{p}}$.
Following \cite{Nakauchi+12}, we assume that either one of the two empirical correlations of LGRBs holds in supermassive collapsar jets: the $E_{\rm{p}}$-$L_{\rm{p}}$ correlation or the $E_{\rm{p}}$-$E_{\gamma,\rm{iso}}$ correlation~\citep{Yonetoku+04, Amati+02}.
If the $E_{\rm{p}}$-$L_{\rm{p}}$ correlation holds for the burst, then the spectral peak energy can be evaluated from the peak luminosity $L_{\rm{p}}$, using the correlation
\begin{eqnarray}
\frac{L_{\rm{p}}}{10^{52}\rm{\,\,erg\,\,s^{-1}}}\simeq2\times10^{-5}\biggl(\frac{E_{\rm{p}}}{1\rm{\,\,keV}}\biggr)^{2.0},
\label{yonetoku}
\end{eqnarray}
as $E_{\rm{p}} = 5.6\times10^2$ and $1.6\times10^2\,\rm{keV}$ for the 1E5 model and the 1E4 model, respectively.
On the other hand, if the $E_{\rm{p}}$-$E_{\gamma,\rm{iso}}$ correlation holds for the burst, $E_{\rm{p}}$ can be evaluated from the isotropic radiated energy $E_{\gamma,\rm{iso}}$, using the correlation
\begin{eqnarray}
\frac{E_{\rm{p}}}{1\rm{\,\,keV}}\simeq80\biggl(\frac{E_{\rm{\gamma,iso}}}{10^{52}\rm{\,\,erg}}\biggr)^{0.57},
\label{amati}
\end{eqnarray}
as $E_{\rm{p}} = 2.6\times10^4$ and $1.4\times10^4\,\rm{keV}$ for the
1E5 model and the 1E4 model, respectively.  We summarize the
observational signatures of the prompt emission from the supermassive
collapsar jets in Table \ref{prompt emission table}, where the
redshift of the burst is set as $z=15$.  In lines 4 and 5, the
spectral peak energy in the observer frame $E_{\rm{p}}^{\rm{obs}}$ is
given by $E_{\rm{p}}^{\rm{obs}}=E_{\rm{p}}/(1+z)$.  From Table
\ref{prompt emission table}, we find that the total energy is much
larger than that of LGRBs, while the peak luminosity is comparable to
them. The accretion time of SMS is much longer than that of LGRB so
that SMS releases much larger amount of energy than WR collapsars
although the luminosity is similar.

\begin{table}[!t]
\begin{center}
\caption{Observational Characteristics of the Prompt Emission at $z=15$
}
\label{prompt emission table}
\begin{tabular}{ccc}
\tableline\tableline
Progenitor Model & 1E5 & 1E4 \\
\tableline
$E_{\gamma,\rm{iso}}$ [erg] & $2.5\times10^{56}$ & $8.4\times10^{55}$\\
$L_{\rm{p}}$ [$\rm{erg\,s^{-1}}$] & $6.2\times10^{52}$ & $5.1\times10^{51}$\\
$E_{\rm{p}}^{\rm{obs}}$ [keV] & $3.5\times10$ & $1.0\times10$\\
$E_{\rm{p}}^{\rm{obs}}$ [keV] & $1.6\times10^{3}$ & $8.6\times10^{2}$ \\
\tableline
\end{tabular}
\end{center}
{\bf{Notes.}}  $E_{\rm p}^{\rm obs}$ in line 4 and $E_{\rm p}^{\rm
  obs}$ in line $5$ show the peak energy of the spectrum predicted by
the empirical relations (\ref{yonetoku}) and (\ref{amati}),
respectively.
\end{table}

Finally, we discuss the detectability of the prompt emission from the
supermassive collapsar jet with detectors like the Burst Alert
Telescope~(BAT) on board the {\it Swift}
satellite~\citep{Barthelmy+05}.  BAT covers the energy range from
$E_{\rm{min}} = 15\,{\rm keV}$ to $E_{\rm{max}}=150\,{\rm keV}$.  The
energy flux detected by BAT is given by
\begin{eqnarray}
f_{\rm{sig}}(t_{\gamma,
  \rm{obs}})=F_{\rm{bol}}(t_\gamma)\frac{\int_{E_{\rm{min}}}^{E_{\rm{max}}}EN(E)dE}{\int_0^{\infty}EN(E)dE},
\end{eqnarray}
where $t_\gamma = t - t_{\rm b}$, $t_{\gamma, \rm{obs}}=(1+z)
t_\gamma$, $N(E)$ and $F_{\rm{bol}}(t_\gamma)$ are the time from the
breakout, the time in the observer frame, the photon number spectrum
and the bolometric flux, respectively.  Empirically, we assume that
$N(E)$ is represented by the Band function~\citep{Band+93} with the
spectral indices of $\alpha=-1$ and $\beta=-2.3$~\citep{Kaneko+06}.
The bolometric flux is given by
\begin{eqnarray}
F_{\rm{bol}}(t_{\gamma, \rm{obs}})=\frac{L_{\rm{\gamma,iso}}(t_{\gamma})}{4\pi{d_{\rm{L}}(z)^2}}\rm{\,\,erg\,\,{s}^{-1}\,\,cm^{-2}},
\end{eqnarray}
where $d_{\rm{L}}(z)$ is the luminosity distance.

\begin{figure}[!t]
\centering
\includegraphics[scale=0.3,angle=270]{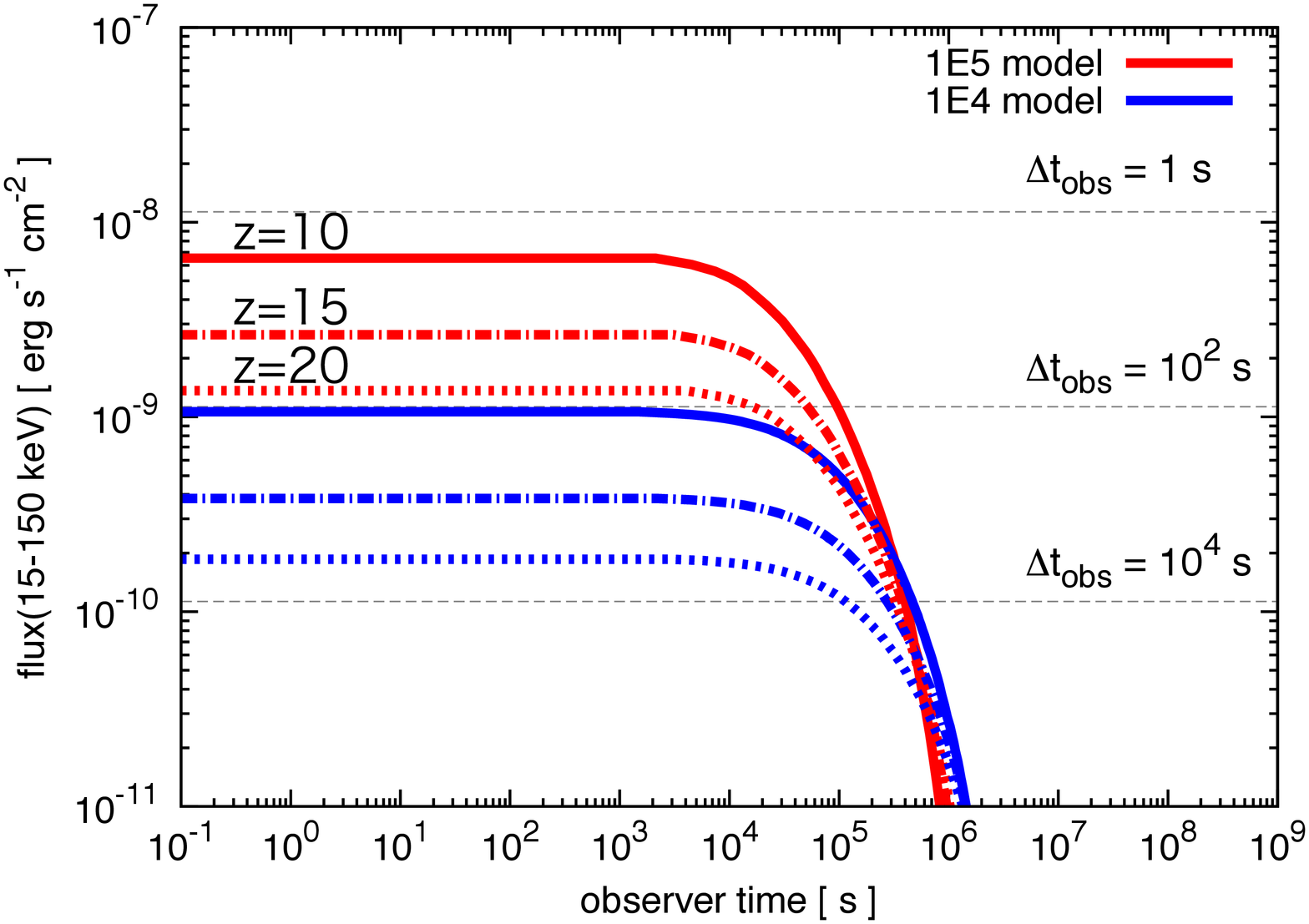}
\caption{ Light curves of the prompt emission of the ULGRBs from
  supermassive collapsars.  The flux is calculated at the {\it Swift}
  BAT energy range~(15-150 keV), assuming that the
  $E_{\rm{p}}$-$L_{\rm{p}}$ correlation holds.  The \emph{red} and
  \emph{blue curves} correspond to the 1E5 and the 1E4 models,
  respectively.  The redshifts of the bursts are $z=10$
  (\emph{solid}), $15$ (\emph{dash-dotted}), $20$ (\emph{dotted}),
  respectively.  The \emph{gray dotted lines} show the BAT
  sensitivities $f_{\rm{sen}}(\Delta t_{\rm{obs}})$ with the
  integration times of $\Delta{t_{\rm{obs}}}=1,$ $10^2,$ $10^4$ s,
  from up to bottom.}
\label{light curve BAT yonetoku image}
\end{figure}

In Figure \ref{light curve BAT yonetoku image}, we show the light
curves of the prompt emission in the case of the
$E_{\rm{p}}$-$L_{\rm{p}}$ correlation.  The red and blue curves
correspond to the 1E5 and the 1E4 models, respectively.  We set the
redshifts of the bursts as $z=10$ (solid), $15$ (dash-dotted), $20$
(dotted), respectively.  The gray dotted lines show the BAT
sensitivities $f_{\rm{sen}}(\Delta t_{\rm{obs}})$ with the integration
times of $\Delta{t_{\rm{obs}}}=1\,$s, $10^2\,$s, and $10^4\,$s, from
top to bottom.  If the burst enters the field of view of the BAT at
some time $t_{\gamma, \rm{obs}}$, and the signal flux is larger than
$f_{\rm{sen}}(\Delta t_{\rm{obs}})$, then it can be observed by BAT up
to $t_{\gamma, \rm{obs}}+\Delta{t_{\rm{obs}}}$.  We can see that the
burst can be detectable up to $z=20$ for
$\Delta{t_{\rm{obs}}}=10^2\,\rm{s}$ in the 1E5 model.  We can also see
that the burst can be detectable up to $z=20$ for
$\Delta{t_{\rm{obs}}}=10^4\,\rm{s}$ in the 1E4 model.

\begin{figure}[!t]
\centering
\includegraphics[scale=0.3,angle=270]{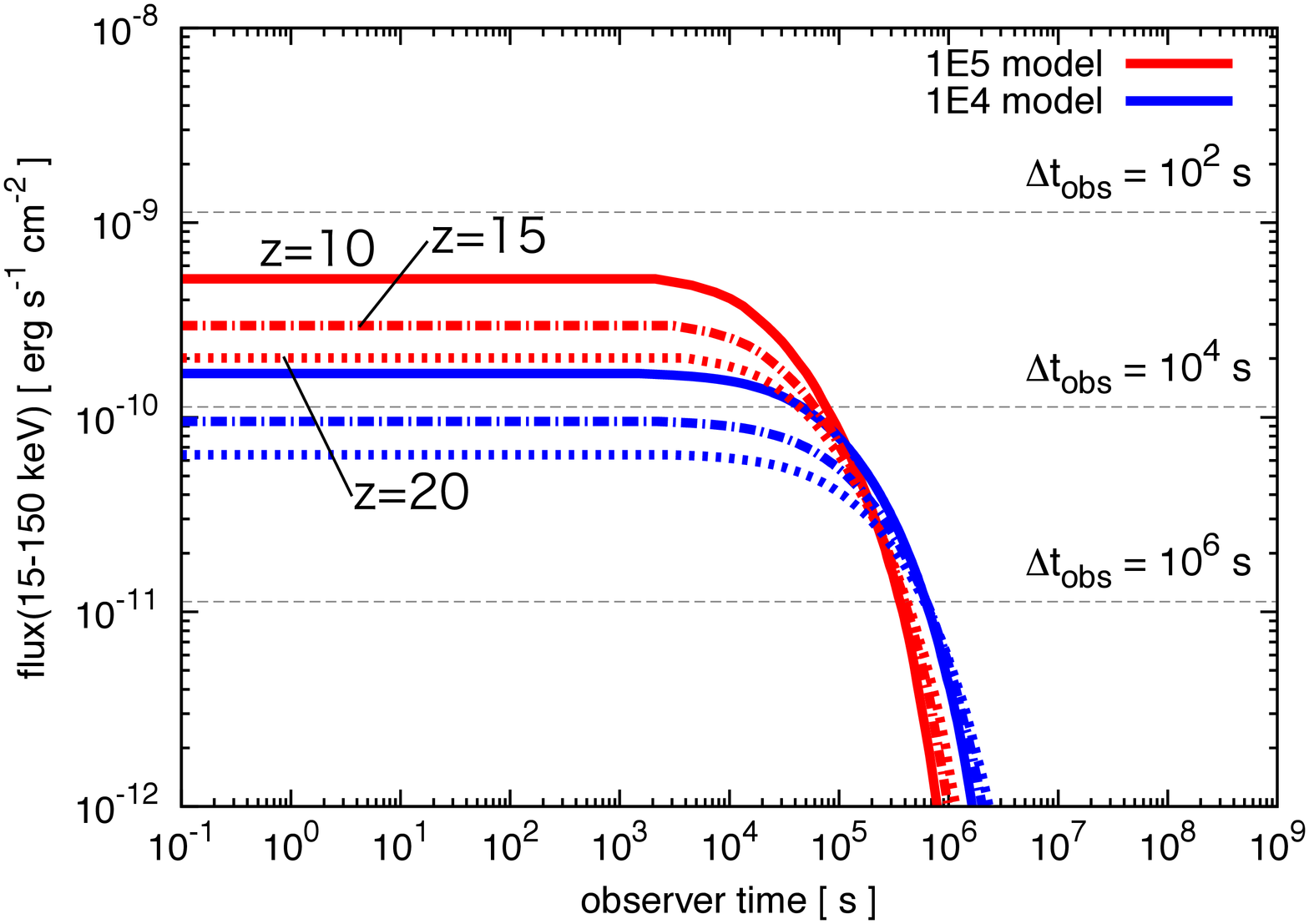}
\caption{The same as Figure \ref{light curve BAT yonetoku image}, but
  the flux is calculated assuming that the
  $E_{\rm{p}}$-$E_{\gamma,\rm{iso}}$ correlation holds.}
\label{light curve BAT amati image}
\end{figure}

In Figure \ref{light curve BAT amati image}, we show the light curves
of the burst in the case of the $E_{\rm{p}}$-$E_{\gamma,\rm{iso}}$
correlation.  We can see that the observed flux is smaller by an order
of magnitude than that of the above case.  This is because the
$E_{\rm{p}}$-$E_{\gamma,\rm{iso}}$ correlation leads to
$E_{\rm{p}}^{\rm{obs}} \simeq 1\,\rm{MeV}$, which is out of the BAT
energy range.  Nonetheless, we can see that the burst can be
detectable up to $z=20$ for $\Delta{t_{\rm{obs}}}=10^4\,\rm{s}$ in the
1E5 model.  On the other hand, in the 1E4 model, longer integration
times~($\Delta{t_{\rm{obs}}} >10^4\,\rm{s}$) are needed to detect the
burst at such a high redshift of 20.

Thus, we conclude that the prompt emission from supermassive collapsar
jets can be detectable as ULGRBs up to $z \sim 20$ even with the
current detectors like BAT.

\section{Discussion}\label{Discussion}

\subsection{Accreting supermassive stars}\label{Accreting supermassive stars}
If the high mass accretion rate of $0.1$-$1\,\rm{\Msun\,yr^{-1}}$
continues, the SMS mass will eventually reach a few times
$\sim10^5\,\rm{\Msun}$ and the GR instability sets in to lead the
formation of a BH and accretion disk system
\footnote{If a SMS rotates, it is stabilized against the GR instability \citep[e.g.,][]{Fowler66,Bisnovatyi-Kogan+67,Baumgarte and Shapiro99}.
However, even if the SMS rotates, it will not acquire more mass than $\sim10^6\,\Msun$ for the highest mass accretion rate of $1\,\Msun\,\rm{s^{-1}}$.
Eventually, it  exhausts the nuclear fuel in $\sim1\,\rm{Myr}$ and collapses to form the central BH.
Actually, the calculation of \cite{Hosokawa+13} shows that the hydrogen burning occurs in the core.}.
Such a high mass
accretion rate onto protostars is obtained if the temperature is high
as shown in Equation (\ref{eq:m_acc}) and second paragraph of Section
\ref{Progenitor Model}.  Recently, \cite{Hosokawa+13} calculated the
evolution of a SMS under such a high mass accretion rate
$\sim0.1$-$1\,\rm{\Msun\,yr^{-1}}$ from the protostar phase to the
final mass of $M_\ast=10^5\,\Msun$.  They found that the SMS evolves
with a very large envelope of $R_\ast \sim 10^{15}\,\rm cm$.  Under
the rapid mass accretion, the stellar envelope has a large opacity
dominated by $\rm{H^-}$ ions, and absorbs heat released by the
contraction of the stellar inner region.  Then, the stellar envelope
expands as the protostar accretes matter \citep{Hosokawa+12}.

\cite{Hosokawa+13}, however, were not able to compute the SMS
evolution beyond $10^5\,\Msun$, because their stellar evolution code
is suffered from the numerical difficulties. The reason is not
clear. If the accretion is stopped before the GR instability sets in,
the final pre-collapse model is similar to the 1E5 model so that the
ULGRB of the SMS is expected as discussed in the previous section. If
the accretion continues, the SMS will enter the GR instability
region. In such a case, we can assume that the SMS has similar density
profile obtained by \cite{Hosokawa+13} when it begins to collapse
through the GR instability.  The structure of the SMS envelope would
not change so much as long as the high mass accretion rate is kept
until the SMS obtains the critical mass.  Therefore, our assumption
may be justified for the envelope, which is important for the
propagation of jet heads as we saw in the previous section.

\begin{figure}[!t]
\centering
\includegraphics[scale=0.3,angle=270]{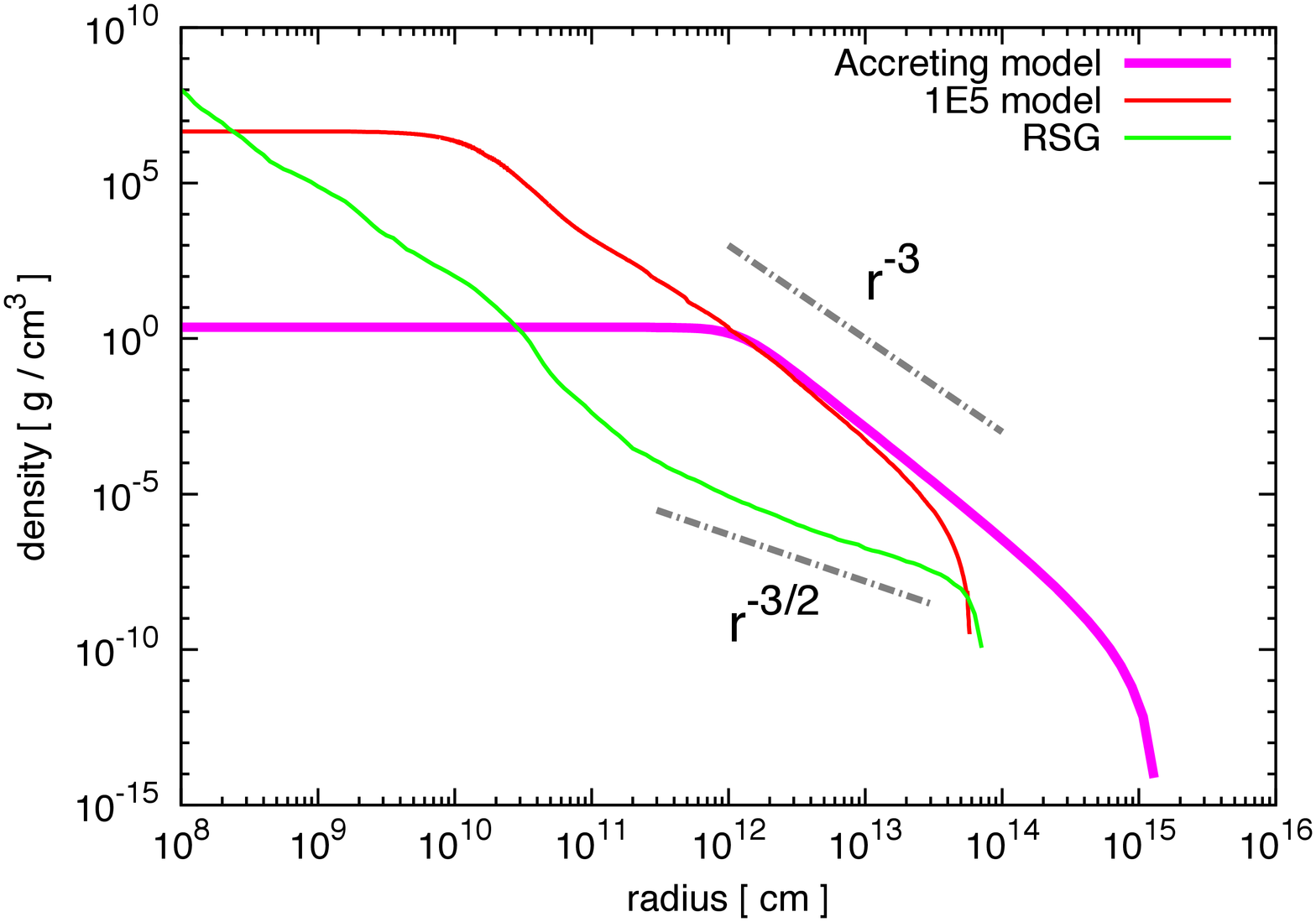}
\caption{Density profile of an accreting SMS with mass of
  $10^5\,\Msun$ (\emph{magenta curve}).  This progenitor also has the
  steep density profile of the envelope (with slope of $\sim3$), where
  a relativistic jet is not decelerated. Note that this progenitor is not
  yet in the pre-collpase phase, but in the accreting protostar phase.
  We can expect that when the GR instability sets in, however, the
  pre-collapse progenitor has the similar density profile especially
  in the envelope as long as the high mass accretion rate is kept.
  With the \emph{red} and \emph{green curves}, we also show the
  density profiles of the 1E5 model and RSG, respectively, for
  references as in Figure~\ref{density profile image}.  }
\label{density profile2 image}
\end{figure}

\begin{figure}[!t]
\centering
\includegraphics[scale=0.3,angle=270]{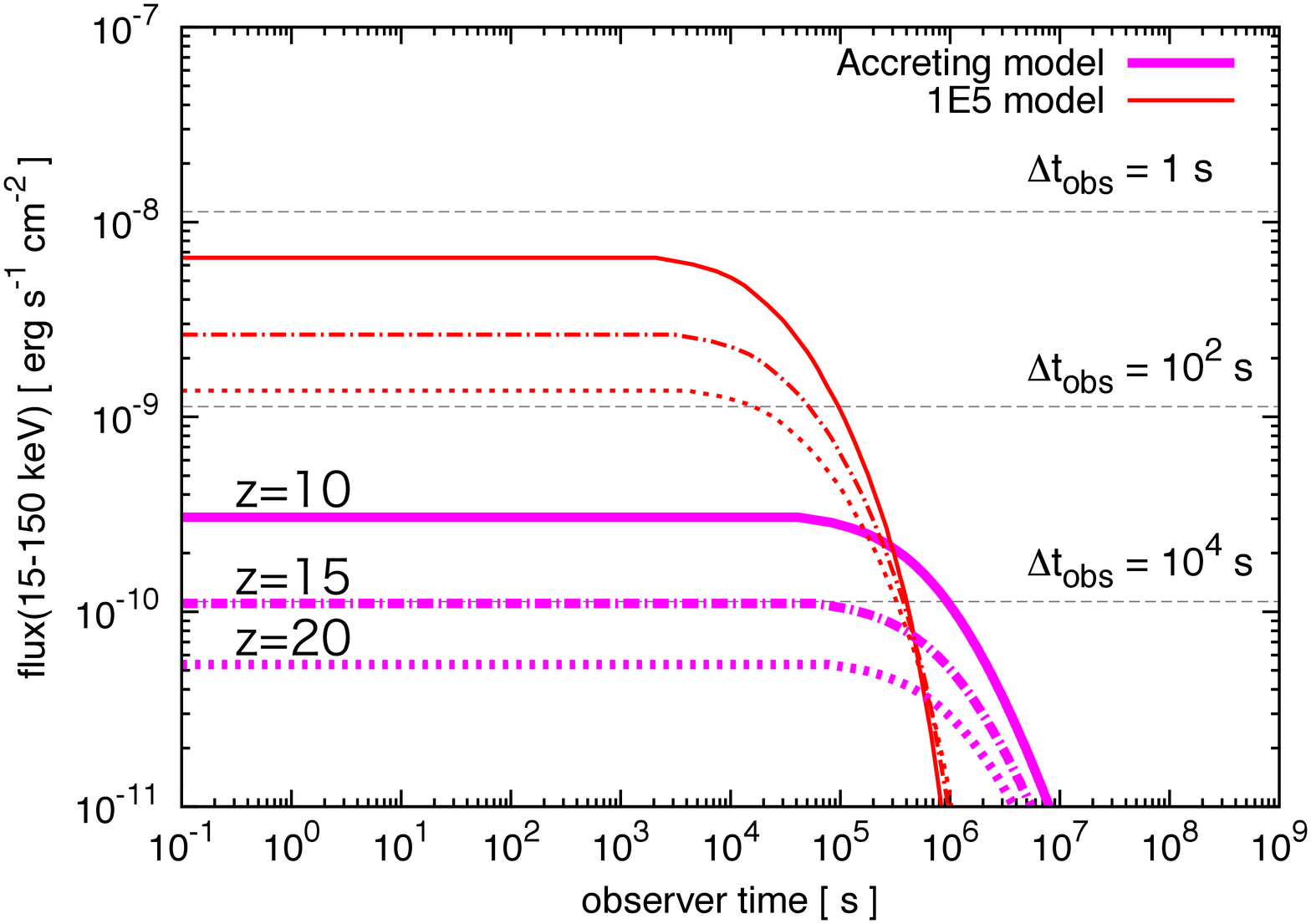}
\caption{ The same as Figure \ref{light curve BAT yonetoku image}, but
  the flux is calculated from the accreting supermassive collapsar
  (\emph{magenta curves}) assuming that the $E_{\rm{p}}$-$L_{\rm{p}}$
  correlation holds.  We also show the light curves of the 1E5 model
  for references.}
\label{light curve BAT yonetoku2 image}
\end{figure}

In Figure~\ref{density profile2 image}, we show the density profile of
the accreting SMS when its mass reaches $M_\ast=10^5\,\Msun$ by using
the magenta curve.  While in \cite{Hosokawa+13}, the density profile
is given as a function of the mass coordinate $M_r$ (in Fig.~3 of
their paper), we show it as a function of radius $r$ by integrating
the mass conservation equation $dr/d M_r = 1/ 4 \pi r^2 \rho(M_r)$.
As shown in Figure 6, the accreting SMS is about $10$ times larger
than a RSG.  Hence one may expect that the jet can not break out the
surface of the star as RSGs.  However, we find this progenitor also has a steep
density profile in the radiation-pressure-dominated envelope.
Therefore, according to the argument in Section \ref{breakout}, we
expect that a relativistic jet is accelerated to break out the surface
of the progenitor star.  We calculate the jet propagation in the same
way described in Section \ref{GRB associated with the SMSs}, and find
that the jet actually breaks out the envelope.  Thus, we expect that
even if the SMS collapses in the accreting protostar phase, it can
produce an energetic explosion.  We also calculate light curves of the
prompt emission, assuming that the $E_{\rm{p}}$-$L_{\rm{p}}$ and the
$E_{\rm{p}}$-$E_{\gamma,\rm{iso}}$ correlations hold.  In Figures
\ref{light curve BAT yonetoku2 image} and \ref{light curve BAT amati2
  image}, we show the light curves.  ULGRBs from accreting SMSs are
dimmer than the GRBs from the 1E5 model.  This is because the
accreting SMSs are larger than SMSs of the 1E5 model.  It takes more
time for a jet to break out the larger envelope.  Thus, when the jet
breaks out the envelope, the accretion rate onto the central BH is
decreased.  The observation time to detect ULGRBs from accreting SMSs
is longer than that of the non-accreting SMSs in Figures \ref{light
  curve BAT yonetoku image} and \ref{light curve BAT amati image}.

\begin{figure}[!t]
\centering
\includegraphics[scale=0.3,angle=270]{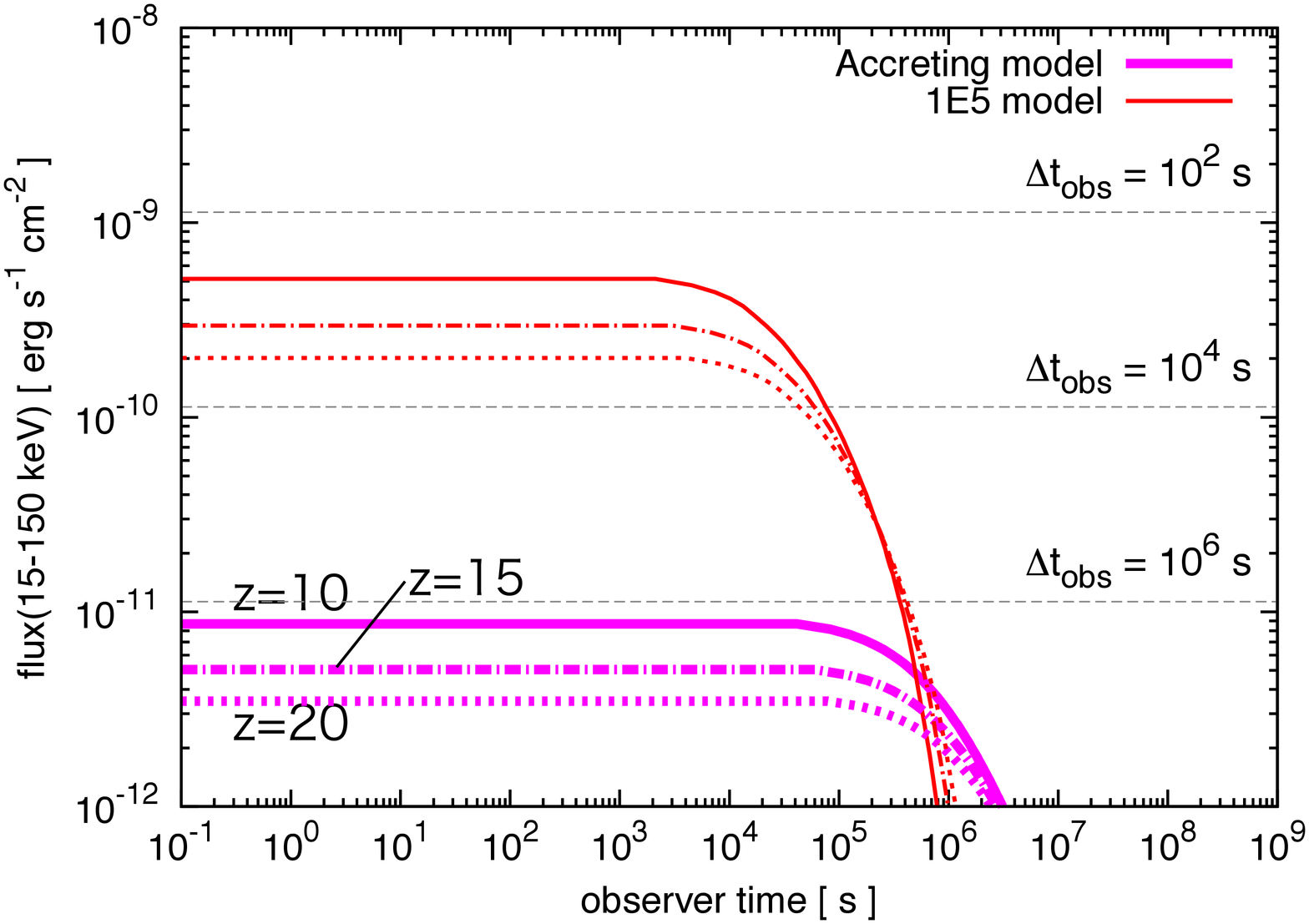}
\caption{The same as Figure \ref{light curve BAT yonetoku2 image}, but
  the flux is calculated assuming that the
  $E_{\rm{p}}$-$E_{\gamma,\rm{iso}}$ correlation holds.  }
\label{light curve BAT amati2 image}
\end{figure}

\subsection{Event Rate}\label{Event Rate}
We briefly discuss the detection rate of the ULGRBs from supermassive
collapsars.  For a given observation time $\Delta{t}_{\rm{obs}}$, the
cumulative number of ULGRBs $\Delta{N(z)}$ which have redshifts less
than $z$ can be calculated from
\begin{eqnarray}
\Delta{N(z)} = \int_0^{z}\Psi_{\rm{GRB}}(z^\prime)4\pi{c}r(z^\prime)^2\biggl|\frac{dt}{dz}\biggl|dz^\prime \Delta{t_{\rm{obs}}},
\label{event rate}
\end{eqnarray}
where $\Psi_{\rm{GRB}}$ is the intrinsic event rate of ULGRBs and
$r(z)$ is the comoving distance to the redshift $z$.  While the
intrinsic event rate $\Psi_{\rm{GRB}}$ is still uncertain, we can
roughly estimate it by using the formation rate of SMSs or DCBHs,
which are theoretically studied in the previous
studies~\citep{Agarwal+12,Dijkstra+14,Yue+14}.

\cite{Yue+14} studied the formation rate of DCBHs in the early
Universe.  They found that DCBHs can be formed from $z=20$ to
$13$~(corresponding to $\sim 150$ Myr), and that the comoving mass
density of DCBHs can be
$\rho_{\rm{DCBH}}\sim2\times10^6\,\Msun\,\rm{Mpc^{-3}}$.  Since they
assumed the typical mass of a DCBH as $M_{\rm{DCBH}}\sim10^6\,\Msun$,
the comoving number density of DCBHs can be evaluated as
$n_{\rm{DCBH}} \sim
\rho_{\rm{DCBH}}/M_{\rm{DCBH}}\sim2\,\rm{Mpc^{-3\,}}$.\footnote{This
  number density seems larger than that of the observed typical
  galaxies.  In \cite{Yue+14}, however, they discussed that only a
  fraction of DCBHs can grow up by the mass accretion and that most
  DCBHs do not acquire sufficient mass and escape our observation.}
Then, we can obtain the intrinsic rate of ULGRBs as
$\Psi_{\rm{GRB}}\sim2\,\rm{Mpc^{-3}}/150\,\rm{Myr}\sim10^{-8}\,\rm{yr^{-1}}\,Mpc^{-3}$.
It should be noted that in this rough estimate, we assume that all the
SMSs collapse to DCBHs after they contribute to ULGRBs.  This rate may
be optimistic.

Substituting the above value into Equation (\ref{event rate}), the
event rate of the ULGRBs on the whole sky can be obtained as
\begin{eqnarray}
\frac{\Delta{N}}{\Delta{t_{\rm{obs}}}}&=&\int_{z=13}^{z=20}\Psi_{\rm{GRB}}(z^\prime)4\pi{c}r(z^\prime)^2\biggl|\frac{dt}{dz}\biggl|dz^\prime\nonumber\\ &\sim&6\times10^2\biggl(\frac{\Psi_{\rm{GRB}}}{10^{-8}\,\rm{yr^{-1}}\,Mpc^{-3}}\biggl)\,\,\rm{yr^{-1}}.
\label{event rate2}
\end{eqnarray}
The detection rate of the ULGRBs is reduced by the beaming factor,
$\Omega_{\rm{beam}}:=\theta^2/2\simeq3.8\times10^{-3}(\theta/5^{\circ})^2$,
since the off-axis bursts are not detectable.  By multiplying the
beaming factor to Equation (\ref{event rate2}), the detection rate is
about one event per year.

The emission from an expanding cocoon fireball might play a key role
to raise the detection rate of the event.  After the jet breakout, the
cocoon also breaks out the star and evolves like a non-relativistic
fireball outside the star~\citep{Kashiyama+13,Nakauchi+13}.  The
cocoon emission will be isotropic and free from the beaming effect.

\subsection{Feedback Effects on the Surrounding Environments}
Various feedback effects are expected from the supermassive
collapsars, since they release a huge amount of energy.  In fact, the
total energy of the cocoon fireball discussed above could be as large
as $E_{\rm{c}}\sim10^{55}$-$10^{56}\,\rm{erg}$, which can be
calculated from Equation \eqref{cocoon energy},
$E_{\rm{c}}=E_{\rm{c}}(t=t_{\rm b})$.  Then, the emission from the
cocoon fireball might be observed as the most energetic supernova
explosions in the Universe.  Moreover, such a violent explosion could
disrupt the host halo, and hinder the remnant massive BHs from growing
up to SMBHs within $\lesssim$ $1\,$Gyr after the BH formation.

In addition, if heavy elements are produced in the jet head and
cocoon, they could contribute to the chemical enrichment of the host
halo.  The metal polluted gas will induce the formation of second
generation of stars.  The line features in the cocoon emission could
also tell us the abundance pattern of the nucleosynthesis to confirm
the SMS origin, although the line may be broad due to the high
expansion velocity.

Recently, \cite{Johnson+13} and \cite{Whalen+13} considered a very
energetic supernova explosion of $\sim10^{55}\,\rm{erg}$ in the first
galaxies, and calculated the dynamical evolution of the blast wave
within the host.  They found that whereas the blast wave engulfs the
entire galaxy, most of its energy is radiated away via efficient
cooling processes, so that the swept up matters~($\sim10^{7}\,\Msun$)
could recollapse to the host $\sim 70\,\rm{Myr}$ after the explosion.
Thus, such an energetic explosion might not hinder the remnant massive
BH from becoming supermassive within $\lesssim$ 1Gyr after its
formation.  On the other hand, the momentum conservation suggests that
the host galaxy is expelled if the explosion energy is larger than
$\sim 10^{56} (M_{\rm halo}/10^7\,\Msun) (v_{\rm esc}/10\,{\rm
  km}\,{\rm s}^{-1}) (\beta_c/0.3)$ erg.  Thus more detailed
calculations are worth while.

Even after contributing to the prompt emission, the relativistic jet
has a huge amount of kinetic energy $E_{\rm{k, iso}} = E_{\gamma,
  \rm{iso}} (1-\epsilon_\gamma) / \epsilon_\gamma \sim
10^{57}\,\rm{erg}$.  This can leads to bright afterglow emissions at
various wavelengths~\citep{Toma+2011,Ioka:2004zk}.  The detection of
such afterglow emissions could provide us rich information about the
surrounding environments of SMSs such as the density and the chemical
composition of the circumstellar medium.

\section{Summary and Conclusions}\label{Summary}
We investigated whether in the early Universe SMSs are able to produce
GRBs according to the collapsar scenario.  Since SMSs have radii at
least as large as RSGs, naively it would seem difficult for a
relativistic to jet reach the surface before the jet engine dies.
Actually calculating the jet propagation in SMSs, however, we find
that jets are able to break out the thick envelope of SMSs.  This is
because the envelope of SMSs is dominated by radiation pressure and
has a steeper density gradient than RSGs in which the gas pressure
dominates.  Our conclusion is that SMSs forming in protogalaxies can
produce violent GRBs.

Based on empirical rules, we find that the collapse of SMSs may be
observed as ULGRBs with a duration of $\sim10^4$-$10^6\,\rm{s}$ by
the current detectors like BAT.  Our optimistic estimates indicate
rates of detectable GRBs of about one event per year.  Comparing with
observations, we can impose some conditions on the intrinsic event
rate which is related to the SMS or DCBH formation rate.  Since GRBs
are collimated bursts, beaming reduces the rate of detectable events.
Therefore, it is important to consider the isotropic emission
accompanying GRBs, e.g., the cocoon fireball emissions.  Studying the
detectability of cocoon fireball emission is an interesting future
work.

The SMS GRBs are very energetic explosions releasing more than
$10^{55}$-$10^{56}\,\rm{erg}$ and sweep up or blow off the matter
in protogalaxies.  As a result, GRBs strongly influence the mass
accretion onto the newborn seed BHs.  This negative feedback needs to
be taken into account when studying the growth of remnant BHs.

\section*{acknowledgments}
We thank T.~Hosokawa, K.~Kashiyama and K.~Inayoshi for fruitful discussions and
comments.  This work is supported in part by the Grant-in-Aid from the
Ministry of Education, Culture, Sports, Science and Technology (MEXT)
of Japan, Nos. 261051 (DN) 24103006, 26287051, 24000004, 26247042 (KI)
24103006, 15H02087(TN).  AH was supported by an Australian Research
Council (ARC) Future Fellowship (FT120100363) and NSF grant
PHY-1430152 (JINA-CEE).

\end{document}